# Generative AI Assistants in Software Development Education

A vision for integrating Generative AI into educational practice, not instinctively defending against it.

**Christopher Bull**; Newcastle University, School of Computing, Newcastle Upon Tyne, UK
**Ahmed Kharrufa**; Newcastle University, School of Computing, Newcastle Upon Tyne, UK

The software development industry is amid another disruptive paradigm change—adopting the use of generative AI (GAI) assistants for programming. Whilst AI is already used in various areas of software engineering [1], GAI technologies, such as GitHub Copilot and ChatGPT, have ignited peoples' imaginations (and fears [2]). It is unclear how the industry will adapt, but the move to integrate these technologies by large software companies, such as Microsoft (GitHub[1], Bing[2]) and Google (Bard[3]), is a clear indication of intent and direction. We performed exploratory interviews with industry professionals to understand current practice and challenges, which we incorporate into our vision of a future of software development education and make some pedagogical recommendations.

*[Sidebar]*

GitHub Copilot example (in Visual Studio Code): Writing the top comment (line 2) in python and pressing enter will use Copilot to suggest a function, which appears dimmed (as "ghost text"). A user can ignore it, browse alternatives, or press Tab to accept the code. The first function in this example is not optimised as it loops through all values up to 'n'. Another comment written later (line 7) requests a more optimised suggestion.

```python
2   # define a function in python that tests
    if a number is a prime number
3   def is_prime(n):
4       if n ≤ 1:
            return False
        for i in range(2, n):
            if (n % i) == 0:
                return False
        return True
5
6
7   # Give a more optimized version of the
    function above
8   def is_prime_optimized(n):
9       if n ≤ 1:
10          return False
11      if n ≤ 3:
12          return True
13      if n % 2 == 0 or n % 3 == 0:
14          return False
15      # if n % 6 == 1 or n % 6 == 5:
16      #     return True
17      i = 5
18      while i * i ≤ n:
19          if n % i == 0 or n % (i + 2) == 0:
20              # print(i)
21              return False
22          i += 6
23      return True
```

It's commonly said that automation and AI will take jobs—this is sometimes true, but it can also create new opportunities; software development is no exception to these concerns [1]. Making aspects of programming easier through GAI may lower the barrier of entry into the profession and allow for developers to shift focus into higher levels of problem solving.

---

[1] GitHub Copilot: https://github.com/features/copilot
[2] Microsoft Bing with ChatGPT: https://blogs.microsoft.com/blog/2023/02/07/reinventing-search-with-a-new-ai-powered-microsoft-bing-and-edge-your-copilot-for-the-web/
[3] Google Bard: https://blog.google/technology/ai/bard-google-ai-search-updates/

## 1.1 Large Language Models in Educational Practice

GitHub Copilot can already successfully do CS1 introductory programming assignments with few or no prompt adjustments [3]–[5]. This has led many to focus on detecting cheating (students using these tools to do the work for them) or adapting assessment methods to make it harder for GAI tools to solve the assignments. This reactionary response is understandable; however, it is contrary to an intention within Computing educational practice to have content and methods informed by industry, and it misses the opportunity to reflect on the inclusion and utilisation of GAI in education.

Software development has seen similar historical shifts, including intelligent code completion (such as Microsoft Visual Studio's IntelliSense[4] and more recently IntelliCode[5]), search engines, and question and answer (Q&A) websites, such as Stack Overflow. Code completion tools sit within an IDE and are a similar, though simpler, concept to GitHub Copilot, providing context aware suggestions to the developer. Search engines are commonly used in everyday development practice, allowing people to search for documentation, code, and solutions. Indeed, there are previous attempts to integrate web search functionality into the IDE to create "example-centric programming" [6]. Easy access to existing solutions and code snippets through these tools affords users the potential to improve productivity and capacity.

Despite this, there are concerns that over-reliance on these tools and sources may negatively impact learners' critical evaluation and problem-solving skills (e.g. copying and pasting code), or the ability to retain information—knowledge obtained through a search engine has lower rates of recall [7]. These all sound familiar to anyone involved in current conversations about the use of GAI, and yet search engines and Q&A websites are readily available and pervasively used. Both approaches attempt to help developers quickly identify solutions, available possibilities, and avoid having to remember functions or their syntax. The criticisms are valid, but the tools and methods have clearly provided an overall benefit.

A fundamental trait of these recent technologies is that they are not truth engines, but language predictors (Large Language Models or LLMs [8])—based on the input, an LLM can predict a response by identifying the next probable series of words (or code). This means it is up to the reader to understand if the generated content is factual and appropriate. Even if these technologies were updated in a way to guarantee factual and correct output (for example, Bing Chat's attempts to cite real sources in its GPT enhanced searches, as opposed to general LLMs potentially making them up), they are still bound by the request of the user—generated responses may be what the user asked for, but not what they intended or require. Knowing what questions to ask (or how to iterate on a question) is a skill.

## 1.2 The Vision

Our vision is to integrate GAI software development tools and practices into programming education, teaching people how and when to use them, like any other tool in their toolkit. GAI technologies are changing how we work, and we believe our graduates need to have adaptable skills to achieve the greatest potential and competitiveness within a changing industry and amongst diverse opinions on the use of GAI tools, maintaining a balance between using the full potential of GAI in coding and the need to retain an objective and critical mindset.

---

[4] IntelliSense: https://learn.microsoft.com/en-us/visualstudio/ide/using-intellisense

[5] IntelliCode: https://visualstudio.microsoft.com/services/intellicode/

Whilst this technology is still developing, and its potential uses and impact are not yet fully understood, it's likely that a programmer's approach to software development will shift, moving the focus from writing typical code to working alongside GAI assistants to design and develop code solutions, as more algorithms are standard enough to be contextualized and generated by GAI. Future programming education may involve learning how to train and fine-tune GAI models, how to refine inputs for these technologies ('prompt engineering' or, more appropriately, 'prompt crafting'), as well as how to integrate them into larger systems and workflows. Additionally, programmers may need to learn how to evaluate and verify the outputs of AI assistants to ensure that they are accurate and dependable. This is not a new required skill, as developers spend more time maintaining existing code than generating new code [9].

We believe shifting the focus of conversations around GAI to the potential it provides and the adaptability of developers working with it will improve the foundation of skills provided during early stages of a professional's education.

## 2 How do Professionals use GAI?

To understand the requirements of teaching with GAI, we set out to understand how professional developers use these tools. We performed some exploratory interviews (during Q1 2023) with industry professionals familiar with GAI tools, asking how they currently use them and what issues they have experienced.

We recruited 5 professional developers (P1-P5) with a range of knowledge and experience (years of professional experience: 1-25, average 11.4), who are familiar with Copilot/ChatGPT:

- P1, P3, and P5 use them regularly in practice.
- P1-P3 use them for personal projects—their use for work is limited by employer restrictions on using GAI, though they have used them professionally for brainstorming ideas, but not generating final code.
- P4 trialled them but did not continue using them.
- P5 uses them for PhD software development and contract work.
- There was a range of enthusiasm for GAI, from enthusiastic (P1, P3, P5) to indifferent (P2, P4).
- P1 and P3 work in some of the world's largest international technology companies, with 25- and 14-years professional experience, respectively.

For recruitment, we shared a call for taking part in interviews using Twitter and LinkedIn as well as directly contacted professionals in our networks. The semi-structured interviews were carried out using Zoom and lasted 45-60 minutes. Consent was given to record the interviews. Deductive thematic analysis was carried out based on our questions around tool use, training/education, management perspective, level of experience, code quality, challenges, limitations, and concerns. 'Desired features' is the only new identified theme. While some points depend on a single participant, we adopted Braun and Clarke's view that "the 'keyness' of a theme is not necessarily dependent on quantifiable measures but rather on whether it captures something important" [10]. However, we do acknowledge the limitation of our small sample size and that this should be taken into consideration when interpreting these results.

## 2.1 Tool Usage

Our participants shared a variety of usage scenarios and justifications. P1 uses ChatGPT primarily, describing it as "the next leap in software development" and considers it "really empowering". The two main types of activity P1 uses it for are "boring stuff, not complex stuff, or just like write [the] first version of anything you want" and for treating it like a "consultant that you brainstorm with". This specifically included repetitive code, infrastructure, "boring" code, system design, brainstorming ideas, and "writing test cases and automation tests". P1 also added that they "mostly use it for system design" and that it "feels like you are sitting with a real person giving you lots of ideas".

> **"This is like the first time I really feel the power of AI" (P1, Expert Software Developer)**

P3 and P5 use Copilot for creating simple functions and P3 uses ChatGPT (like P1) for "more sophisticated stuff". P3 anecdotally told of another developer who joked about people who use these tools (that they are not good programmers), but now uses the tools themselves.

P2 discussed their use of ChatGPT and how they appreciate the available context which allows them to iterate questions. Though P2 also considers these tools as an entry point for learning, but not for daily work, differing from all the other participants. P2 stated that they do not use Copilot or other tools, like IntelliSense, "so you actually have to actively sort of remember them [e.g., function names]". However, they did add that Copilot can show you new ways of doing something.

P2 and P3 expect no significant implications on industry but think they may improve productivity and speed up coding as "it will write a lot of code for you automatically" (P3). However, P2 and P3 mentioned that software development is more than just coding.

## 2.2 Copilot vs ChatGPT

Whilst Copilot and ChatGPT are different tools that can serve different purposes, we asked our participants about how they used the tools differently. P1 prefers the conversational aspect to ChatGPT, highlighting that Copilot cannot give feedback, maintain contextual awareness, or have access to a historical chat. Another big advantage for P1 is that when ChatGPT "writes the code, it explains the code". Copilot, on the other hand, looks at your code but does not understand the world. ChatGPT better understands the world, so you can link what you are building to the world (e.g., scraping data from known platforms); it has the internet as a context because of the way it was trained (P1).

P2 and P3 referred to another key advantage of ChatGPT: As it does not have access to the code, one can abstract a work-related problem to be a general one and ask it to ChatGPT, but this is not the typical use case with Copilot. This inherently has benefits for critically thinking about a problem and translating that into a different context.

Whilst the description of the GAI features used was expected by the authors, the preference and prevalence of praise for the conversational aspects of ChatGPT were surprising.

## 2.3 Training and education

All participants agree that there is no need to train developers specifically on the use of these tools; a brief demonstration of features is sufficient. However, they did all also agree that to benefit from these tools, people should have a grasp of programming fundamentals or a curious nature. P3 and P5 feel that it is about asking the right questions, understanding their capabilities, and not overestimating them. P2 goes further to say that Copilot is just another IDE tool like autocomplete.

P2 and P5 think the training should be on the fundamentals of programming to be able to judge correct from incorrect code and the most efficient code from the inefficient one. From a formal education perspective, P3 expressed concerns about possible overreliance on the tools by students and that they may end up not doing as much coding themselves, emphasizing that businesses and education both have to adapt because 'the use of these tools is inevitable'.

## 2.4 Junior vs Expert Users

P1, P2, and P3 think the tools benefit everyone, helping each differently depending on their level of experience. P1: For novices, they are going to help write basic lines of code and explain them, and for experts they can help at system design level. P3 and P4 felt that there is a possibility that for junior developers they can "do more harm than good because they don't get to learn the basics" (P3) but can also see how they can benefit them as well, and they benefit both junior and expert developers worrying about the bigger picture. P2 also joked about how such tools can potentially slow down an expert programmer: a good solution requires a good prompt, and sometimes it is faster to write the solution than figuring out the good prompt.

P5 had an interesting discussion about not considering people junior or senior developers. As with many areas of software development, beyond fundamentals, most programmers with an expertise will be considered a novice in some other contexts. Therefore, GAI tools may be used by senior experts who are working in an unfamiliar language or exploring application domains or SDKs that are new to them.

## 2.5 Code Quality

This was discussed on a few occasions, both in terms of code produced through a GAI tool, but also the effects that GAI has on the development processes.

P1 used GAI to generate unit tests and assist with refactoring but insisted that "you always need a human to look at [the code]". Similarly, P2 mentioned that one can use their generated code as a starting point, but they need a human to judge the quality of work – something already practiced by many organizations for human generated code as mentioned by P3 who stated that in their organization code is reviewed by two others before getting approved.

With a balanced viewpoint, P3 discussed how GAI tools can improve code quality by generating documentation, code comments, and possibly good quality code. However, with inadequate QA, P2 and P3 feel the code could introduce bugs through poor code suggestions.

## 2.6 Limitations

All participants referenced the risk of a GAI tool giving an incorrect but confident response. Sometimes answers can look "so convincing that basically you think all this should work, but it didn't" (P2).

Beyond this, P1 and P2 raised the concern that the model was trained on data up to a point and may not use the latest versions of libraries or APIs. Copilot is also said to work best for generating small pieces of code (P3).

P2's main criticism of Copilot is that "it doesn't really have the context of what you are working on". They are not referring to ChatGPT-like conversational context, but to the bigger context of the project and its goals, related files, etc.

### 2.7 Desired Features

As an IDE integrated tool, according to P1, it would be good if Copilot had access to files in the project whilst also understanding its file structure and business logic.

P1 also had other requested features, such as being able to run services offline, linking to external resources (citing sources), and providing diagrams to explain software design.

Another requested Copilot feature was for it to be able to crawl your code and make suggestions to optimise your code.

## 3 Pedagogical Recommendations

From the insights gained through the interviews, we envisage several techniques that could help support the inclusion of GAI into programming education:

### 3.1 Scaffolding and Fading

Participants (P1-P3,P5) talked about the different levels of support that juniors and experts may need from GAI and how an expert in one topic is a junior in another. From a pedagogical perspective, GAI tools can be used to provide effective scaffolding and fading to support the student (or junior developers) in their learning and to offload some of that scaffolding burden from the educator. Scaffolding and fading [11] builds strongly on Vygotsky's Zone of Proximal Development (ZPD) [12] which refers to the difference between what a student can achieve alone and what they can achieve under guidance. In this context, scaffolding refers to the reduction of the load on the teacher in terms of maintaining the student's interest and focus on the task by keeping the task difficulty within the student's ZPD by providing a structure to the task and demonstrating how to achieve the required goals to control frustration when faced with complex tasks. Fading, on the other hand, refers to the progressive reduction of such support to the point where the student can solve the problem alone.

Sweller's work on worked examples in cognitive load theory [13] provides a more concrete method of implementing scaffolding and fading. Worked examples provide scaffolding by providing learners with a demonstration of how a problem is solved to complete the task using representative examples with a clear step by step explanation of the process (very similar to what ChatGPT already does, or the comments inserted by Copilot). The aim is to help the learner develop their problem-solving skills and deepen their understanding of the covered concept by providing a clear model to follow. Sweller argues that this approach can help reduce the cognitive load on the learners and help them focus on the key concepts of the problem. Passive learning through glossing over such examples can be avoided by providing partial worked examples with the learner having to complete some key aspects or adopting a paired approach (study an example to solve a similar problem).

GAI tools can provide such pedagogical support to learners if integrated within IDEs with these approaches in mind. If fine-tuned on model code and coding patterns by the educational institutions, many of the concerns around code quality and teaching incorrect practises can be addressed. A speculative feature could also be a variable dial or sliders provided within an IDE (or backend service) to adjust how much a GAI tool can scaffold or fade within an IDE's runtime, for example limiting the number of requests.

### 3.2 Assessment Changes

While P2 and P3 refer to abstracting their problems before using ChatGPT due to privacy concerns (section 2.2), their comments highlight the fact that it is easier to use such tools on abstract or common problems than having to explain the full context of a specific problem. GAI tools perform impressively well for common tasks, code snippets, and simple algorithms that have been carried out repeatedly. However, when the problems given to students are at higher level, or that link to something unique to them (e.g., part of a bigger project they are working on or explained with reference to their local context – a limitation that P2 referred to), the problems cannot be as easily delegated to GAI tools, or GAI tools may not necessarily provide the best possible solution. As such, we may see a shift in assessment to more contextualized and applied problems as opposed to those that are more generic in nature.

Changing assessments to be more context-specific helps in two ways: (1) it allows students to use GAI tools whilst still engaging in critical problem solving and creative thinking, and (2) it helps reduce the potential for GAI tools to be used for cheating (having the tool generate a full solution for you).

### 3.3 Transitional period

A transitional period for new learners could be considered, in a similar way that junior students are not trained immediately with advanced debugging tools, but instead taught the basics of how to read terminal output and error messages first. P2, P3, and P5 talked about learning the fundamentals first before starting using such tools. We do not have to consider immersing students with these technologies from day one. While students cannot be prevented from using such tools if they wanted to, this can at least be enforced inside a computing lab.

GAI has the potential to create unexpected or unintended outcomes, and junior students may not have the experience or knowledge to understand the implications of using such technologies. Therefore, it may be better for students to focus on developing their knowledge of core computing concepts, before attempting to use GAI. Moreover, including 'code reviews' as a common practice early on will also equip learners with the skills to critically examine code written by others (by peers, from open-source repositories, or given as a generated answer, to eventually apply this skill to GAI generated code). This also prepares students for a common professional practice, where their code is reviewed by peers before being approved, as explained by P3 (section 2.5). These practices will ensure that students have the necessary understanding of the fundamentals, and critical skills to use GAI properly and safely.

## 4 Challenges with GAI

Whilst these technologies have the potential to disrupt the software development industry, we would be remiss to not acknowledge some fundamental challenges with these technologies:

- **Copyright**–GitHub Copilot uses OpenAI's Codex, which is trained on open-source software projects. There are concerns about copyright (the model ignores projects' licenses). The argument against this is that Copilot does not use code snippets for a user to copy and paste, but instead is a predicted series of words.
- **Bugs and vulnerabilities**—Code generated by a GAI may contain bugs if the sources contained bugs. A common way to write some code is not necessarily the correct way to write some code.

- **Convincing but incorrect code**—Code can be generated that is functionally correct and compiles, but it may not have the desired or requested outcome.

We are optimistic that these issues can be overcome through permissive sourcing for modelling, or additions of built-in QA. Though some challenges are far more difficult to overcome and are active areas of research:

- **Sustainability**—The computational cost of using and training these GAI tools is high. There is a high energy requirement and AI generally "may have profound implications for the carbon footprint of the ICT sector" [14].
- **Bias**—AI systems can exhibit or perpetuate numerous biases. Biases need to be better understood, mitigated against (in the training data, system designs, and more), and better accounted for [15].

## 5  Creating Problem Solvers

GAI tools offer an exciting potential for software development education that could help learners of all levels to become more efficient and effective software developers. By incorporating GAI into educational practices, educators and students could benefit from the capacity of GAI to reduce the time taken to complete tasks, or to allow for more sophisticated exploration and experimentation.

Our pedagogical suggestions are the beginning of what we consider to be future tool support for GAI. Ultimately, we want to be graduating future problem solvers.